\documentclass[british]{elsarticle}
\usepackage[T1]{fontenc}
\usepackage[latin9]{inputenc}
\usepackage{geometry}
\usepackage{parskip}
\usepackage{color}
\usepackage{verbatim}
\usepackage{float}
\usepackage{url}
\usepackage{amsbsy}
\usepackage{graphicx}

\makeatletter

\providecommand{\tabularnewline}{\\}

\makeatother

\usepackage{babel}
\begin{document}
\begin{frontmatter}
\title{Alpha Background in Multi-Grid Neutron Detectors}
\author[ess]{A. Backis}
\author[ess]{C.-C. Lai}
\author[gla]{J.R.M. Annand\label{cauth}}
\ead{john.annand@glasgow.ac.uk}
\author[ess]{K.G. Fissum}
\author[Kra]{G. Zuzel}
\author[Kra]{M. Czubak}
\author[gla]{K. Livingston}
\address[ess]{European Spallation Source ERIC, SE-221 00 Lund, Sweden}
\address[Kra]{M. Smoluchowski Institute of Physics, Jagiellonian University, Lojasiewicza
11, 30-348 Kraków, Poland,}
\address[gla]{School of Physics and Astronomy, University of Glasgow G12 8QQ, Scotland,
UK}
\cortext[cauth]{Corresponding Author}
\begin{abstract}
Alpha emission from actinide impurities in Al is a source of background
counting rate in Multi-Grid type detectors of thermal neutrons. The
alpha emission rates from samples of radio-purity Al and $\mathrm{Al/B_{4}C}$
composite, used in grid construction, were measured on a large-area,
low background spectrometer. Although the alpha emission rate from
the composite was a factor $\sim280$ higher than radio-pure Al, $\mathrm{25\:\mu m}$
Ni plating of the composite reduced the rate by a factor $\sim1170$.
Background counting rates in two Multi-Grid prototypes were compared.
They used identical configurations of $\mathrm{B_{4}C}$-coated, radio-pure
Al normal blades for the grids, but the first employed radio-purity
Al for the radial blades, while the second used Ni-plated $\mathrm{Al/B_{4}C}$
on the radial blades. The background rate from the second prototype
was around 20\% of that from the first.
\end{abstract}
\end{frontmatter}

\section{Introduction}

\begin{figure}[H]
\includegraphics[width=1\columnwidth]{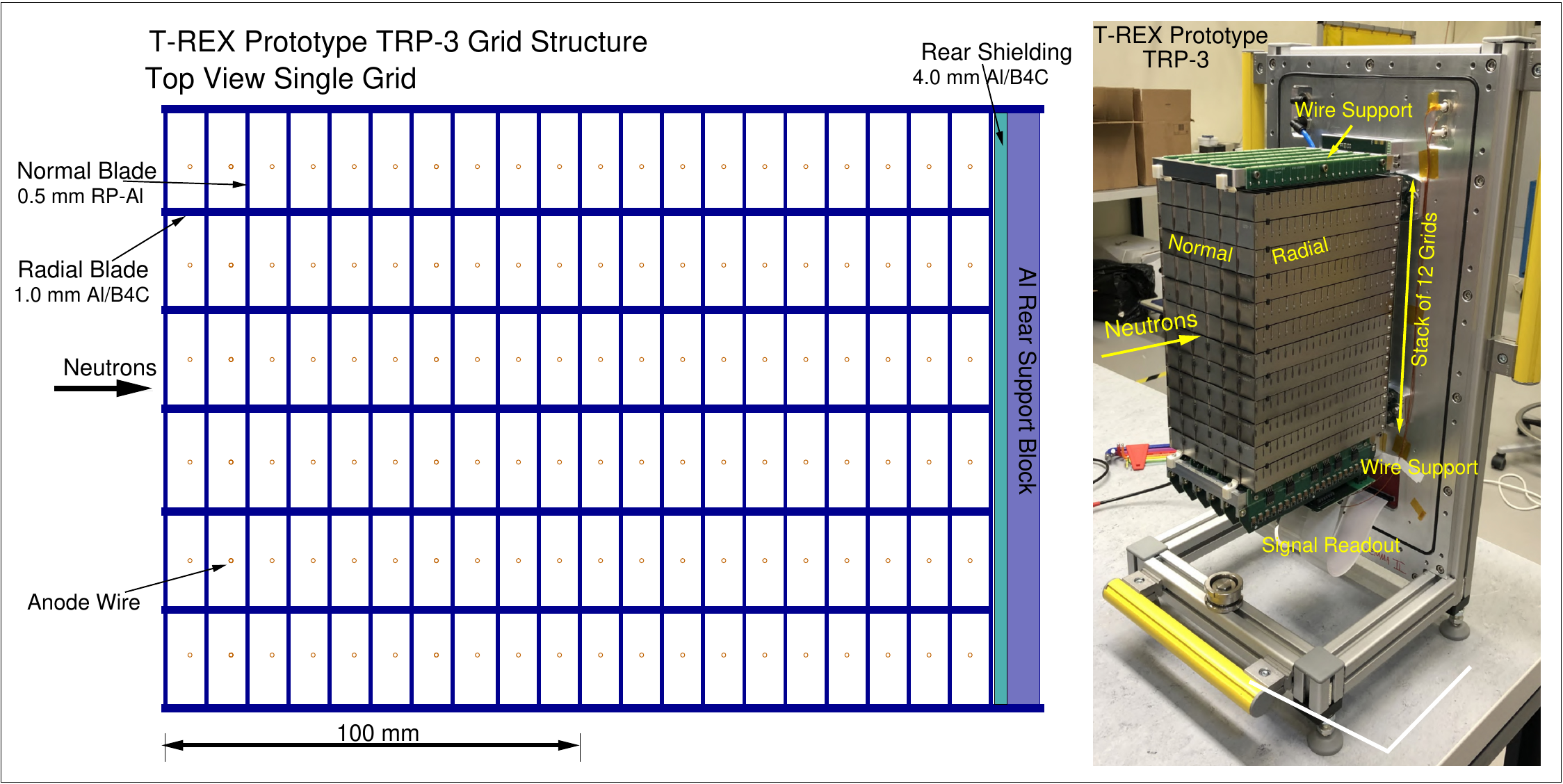}

\caption{\protect\label{fig:Multigrid}Left: prototype T-REX grid. Each voxel
has an internal dimension of 22.5~mm (x) by 24.0~mm (y) by 0.95~mm
(z). Right: photograph of the 12-grid T-REX prototype TRP-3, with
the outer canister removed to show the multi-grid structure.}
\end{figure}

Alpha emission from trace actinide impurities occurring in Al is a
source of background \citep{Birch} in gaseous detectors of thermal
neutrons, which employ neutron capture on $\mathrm{^{10}B}$, to produce
detectable ions. Al is the main structural material in most neutron
spectrometers of this type, due to its relatively low neutron scattering
cross section. Here alpha background has been investigated for prototypes
TRP-1 and TRP-3 of Multi-Grid \citep{ILL} columns for the T-REX spectrometer
at the European Spallation Source (ESS) \citep{ESS}.

The Multi-Grids are stacks of grids, each a rectangular lattice of
normal and radial Al blades (Fig.\ref{fig:Multigrid}). The grids
form the cathodes of a voxelised proportional counter (VPC), with
wires strung through the centres of each grid voxel providing the
anodes. The normal blades are coated with $\mathrm{^{10}B}$-enriched
$\mathrm{B_{4}C}$, and neutron capture in the $\mathrm{B_{4}C}$
film produces $\mathrm{^{4}He}$ and $\mathrm{^{7}Li}$ ions, one
of which escapes into the VPC gas (e.g. Ar-CO2) giving a detectable
signal. Al is the main structural material for the grids and the actinide
impurities will generally be the long-lifetime $\mathrm{^{238}U}$
and $\mathrm{^{232}Th}$ isotopes and their daughters. The alpha emission
spectra span energies up to $\sim9$~MeV, as opposed to $\mathrm{\sim1.4~MeV}$
for post-neutron-capture ions, so this background cannot be rejected
by pulse height (PH) discrimination.

Radio-purity Al (RP-Al) is employed on the normal MG blades to minimise
actinide background. The $\mathrm{B_{4}C}$ coating of 1-2 um allows
the ions from neutron capture to escape into the chamber gas, but
is too thin to stop most actinide alphas. Both prototypes employ this
pattern of normal blades. 

The radial blades of prototype TRP-1 are uncoated RP-Al, while prototype
TRP-3 has radial blades of Ni-plated $\mathrm{Al/B_{4}C}$ composite.
The composite provides internal shielding against neutron multiple
scattering inside the spectrometer, while the Ni plating has been
added to suppress alpha emission, as was observed in Ref.~\citep{Birch}.

Measurements of alpha emission rates from five samples of MG blade
material are presented in Sec.~\ref{sec:alpha=000020activity} and
simulations of the alpha stopping power of Ni plating described in
Sec.~\ref{sec:Modeling}. Sec.~\ref{background} reports on background
rate tests performed at ESS, and conclusions are summarised in Sec.~\ref{subsec:Conclusions}.

\section{\protect\label{sec:alpha=000020activity}Alpha activity measurements}

\begin{figure}[H]
\includegraphics[width=1\columnwidth]{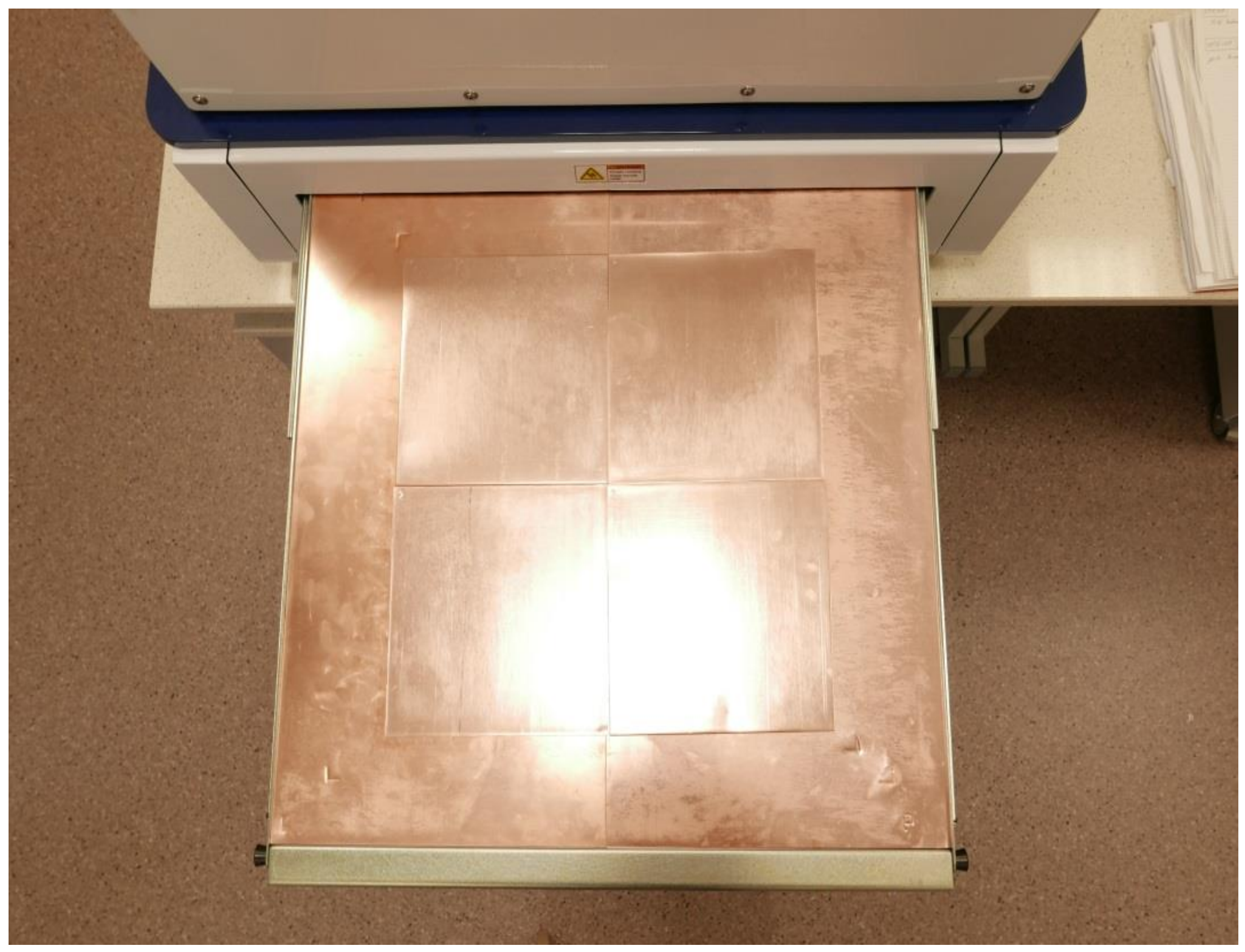}

\caption{\protect\label{fig:Spect}Open tray of the large-surface low-background
alpha spectrometer. The active area is $43\times43\:\mathrm{cm^{2}}$
(the space occupied by four smaller Cu plates on the picture). Samples
(sheets) up to 3 mm thick may be accommodated.}
\end{figure}

Alpha emission rates from five samples, as well as no-sample background,
were measured using an XIA UltraLow-1800 low background alpha spectrometer
(LBAS) \citep{krak0,XiA1800}. The LBAS is an ionisation drift chamber
using continuously-flowed argon as the counting gas. It can measure
sample sheets, up to an area of $43\times43\:\mathrm{cm^{2}}$ and
thickness 3~mm. In order to minimise background counting rates the
sample tray is covered with Oxygen-Free High Thermal Conductivity
(OFHC) copper with etched and electro-polished surfaces. Fig.~\ref{fig:Spect}
shows the open tray with four smaller Cu plates covering the active
area of the detector.

\begin{figure}[H]
\includegraphics[width=1\columnwidth]{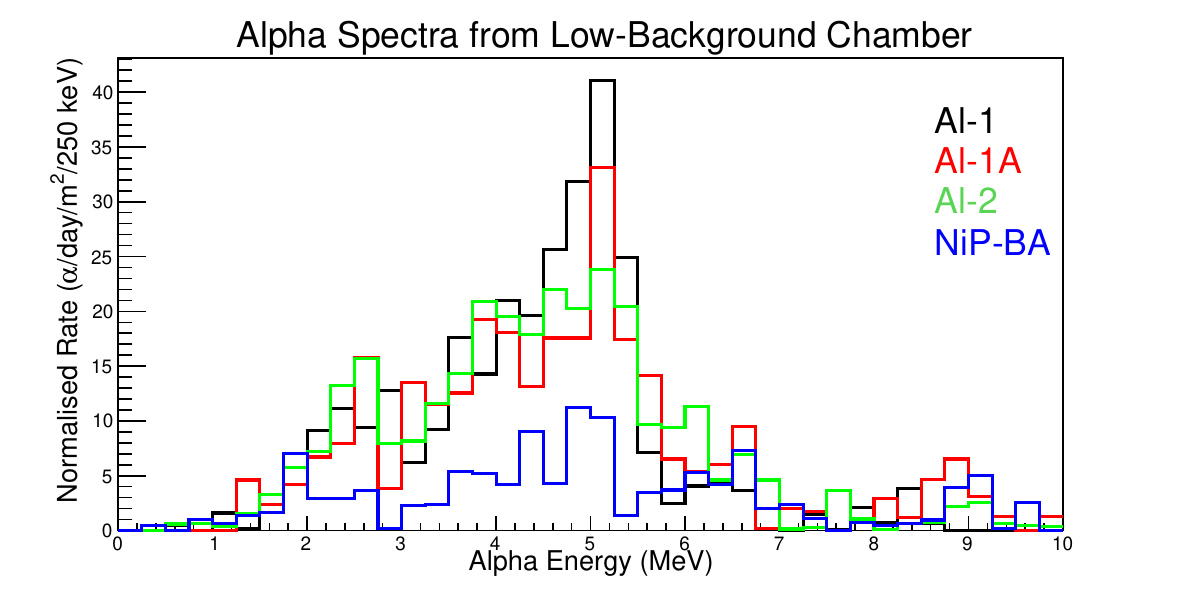}

\caption{\protect\label{fig:Alpha-pulse-height}Background-subtracted alpha
pulse height spectra from sample sheets Al-1, Al-1A, Al-2 and NiP-BA
(Table~\ref{tab:samples})}

\end{figure}

Five sample sheets were tested:
\begin{enumerate}
\item Al-1 0.5~mm radio-purity aluminium (RP-Al).
\item Al-1A 0.5~mm RP-Al, sample 1 after cleaning.
\item Al-2 0.5~mm RP-Al.
\item BA: 1.0~mm $\mathrm{Al/B_{4}C}$ composite, 31\% $\mathrm{B_{4}C}$
by weight
\item NiP-BA: 1.0~mm $\mathrm{Al/B_{4}C}$ composite, electroless plated
\citep{electroless} with a nominal $\mathrm{25\:\mu m}$ of Ni-P
alloy (NiP), approximately 9\% P by weight.
\end{enumerate}
Energy spectra from alpha emission were measured from detector threshold
at 1.5 MeV up to 10.0~MeV. Background was accumulated for 110 days,
while sample-in times were shorter, dependent on the alpha rate. Background-subtracted
pulse height spectra obtained for four of the samples are shown in
Fig.~\ref{fig:Alpha-pulse-height}. The y-scales have been normalised
to give the alpha counting rate per day for a surface area of $\mathrm{1\:m^{2}}$
and an energy bin width of 250~keV.

Table~\ref{tab:samples} gives the dimensions, measurement duration
and alpha surface emission rate for the five samples and a background
measurement with no sample installed, which has been subtracted from
the sample-in rates. Alpha rates from the 2 sheets of RP-Al are very
similar and Al-1 improved marginally, after cleaning with isopropanol
to remove any surface deposits. The other 3 samples (Al-2, BA and
NiP-BA) were also cleaned before measurement took place. 

$\mathrm{Al/B_{4}C}$ composite does not incorporate RP-Al and the
rate from sample BA was $284\pm22$ times higher than Al-2 which was
RP-Al. However the NiP plating reduced rate by a factor $1170\pm330$
compared to BA, so that the NiP-BA rate was $0.24\pm0.07$ that of
Al-2.

\begin{table}[H]
\begin{center}%
\begin{tabular}{|c|c|c|c|}
\hline 
Sample & Size & Time & Surface alpha rate\tabularnewline
\hline 
 & (cm) & (days) & ($\mathrm{\alpha/day/m^{2}}$)\tabularnewline
\hline 
\hline 
Al-1 & $40\times40\times0.05$ & 12 & $320\pm20$\tabularnewline
\hline 
Al-1A & $40\times40\times0.05$ & 11 & $280\pm20$\tabularnewline
\hline 
Al-2 & $40\times40\times0.05$ & 13 & $310\pm20$\tabularnewline
\hline 
BA & $30\times25\times0.10$ & 0.3 & $79440\pm2300$\tabularnewline
\hline 
NiP-BA & $30\times25\times0.15$ & 21 & $68\pm19$\tabularnewline
\hline 
Background & - & 110 & $167\pm6$\tabularnewline
\hline 
\end{tabular}\end{center}

\caption{\protect\label{tab:samples}Parameters of the investigated samples:
dimensions, measurement times and measured surface alpha emission
rates. The sample emission rates are after background subtraction.}
\end{table}

\section{\protect\label{sec:Modeling}Modelling Alpha Emission}

Geant-4 class G4RadioactiveDecay \citep{G4,G4Rad} was used to model
alpha emission from $\mathrm{^{238}U}$ and $\mathrm{^{232}Th}$ isotopes
and their daughter products, embedded in a $\mathrm{400\times400\times0.5\:mm^{3}}$
sheet of Al. G4RadioactiveDecay models the decay sequences using an
evaluated nuclear data file \citep{Ensdf}. The sample sheet was located
in a detector consisting of a $\mathrm{600\times600\times50\:mm^{3}}$
volume of Ar at STP, contained within a thin Al box. Alpha decay chains
were started at coordinates chosen randomly within the volume of the
sample (excluding any coating layer).

\begin{figure}[H]
\includegraphics[width=1\columnwidth]{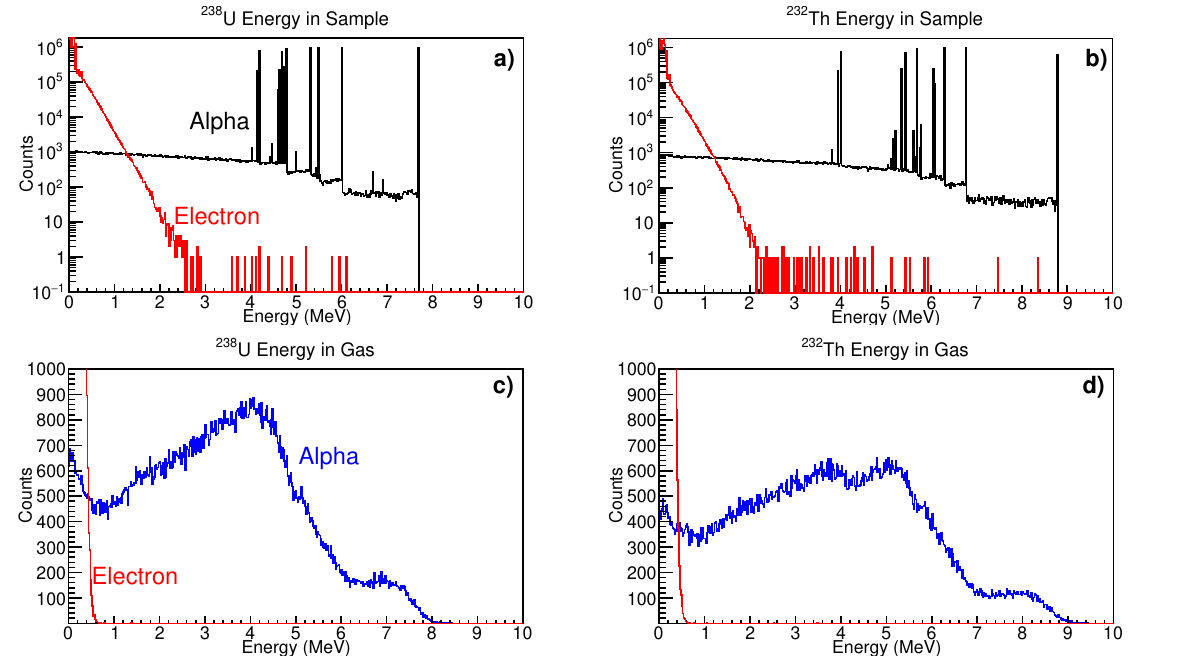}

\caption{\protect\label{fig:Alpha-spectra}Alpha and electron energy deposition
spectra in the sample and detector gas.}

\end{figure}

Spectra of the energy deposited in the sample and detector are displayed
in Fig.\ref{fig:Alpha-spectra}. Alpha particle energy-loss distributions
(black) inside the sample (Fig.~\ref{fig:Alpha-spectra}a,b) show
the emission lines, where the particle has stopped inside the Al,
superimposed on a continuum where the particle has escaped into the
gas. Energy-loss distributions of the electrons produced by $\beta$
decay are also shown in red. 

Corresponding spectra of energy deposition in the detector gas are
shown in Fig.~\ref{fig:Alpha-spectra}~c,d for alphas (blue) and
electrons (red). Particles reaching the gas have lost variable amounts
of energy in tracking through the sample (and any coating) so the
alpha line structure is smeared out. The 1.5~MeV energy threshold
of the measurement is well above the simulated maximum electron energy
deposit in gas.

\begin{figure}[H]
\begin{center}\includegraphics[width=0.5\columnwidth]{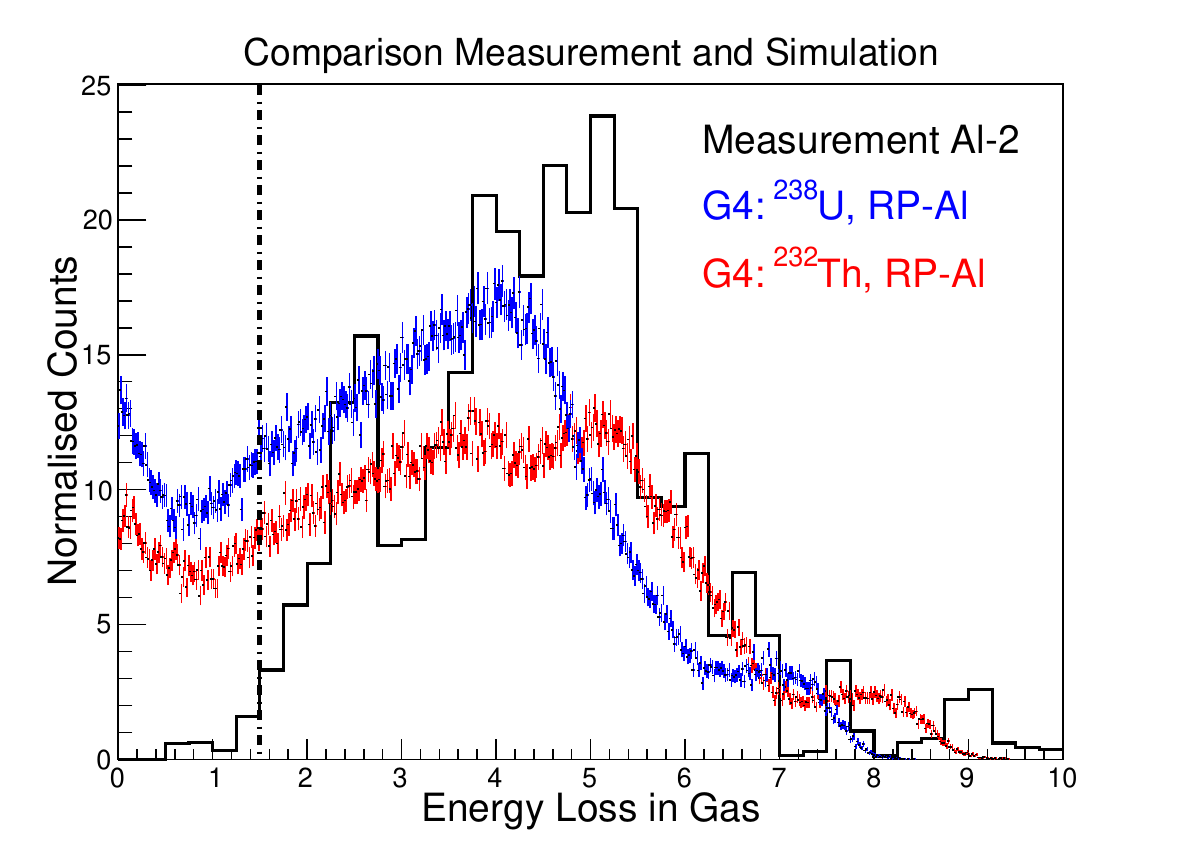}\end{center}

\caption{\protect\label{fig:Comparison-of-alpha}Comparison of alpha spectrum
measured for sample Al-2 with Geant4 simulations of alpha emission
from $\mathrm{^{238}U}$ and $\mathrm{^{232}Th}$ (and their daughters)
embedded in a 0.5~mm thick Al sheet. The Geant4 distributions have
been scaled for comparison.}

\end{figure}

Fig.~\ref{fig:Comparison-of-alpha} compares the simulated pulse-height
distributions in Ar gas with the measurement of sample Al-2, where
the simulated distributions have been normalised to bring them on
scale with the measurement. The simulated distributions are qualitatively
similar in shape to the measurement, but the low counting rate and
limited statistical precision of the measurement limit any quantitative
comparison.

\begin{figure}[H]
\includegraphics[width=1\columnwidth]{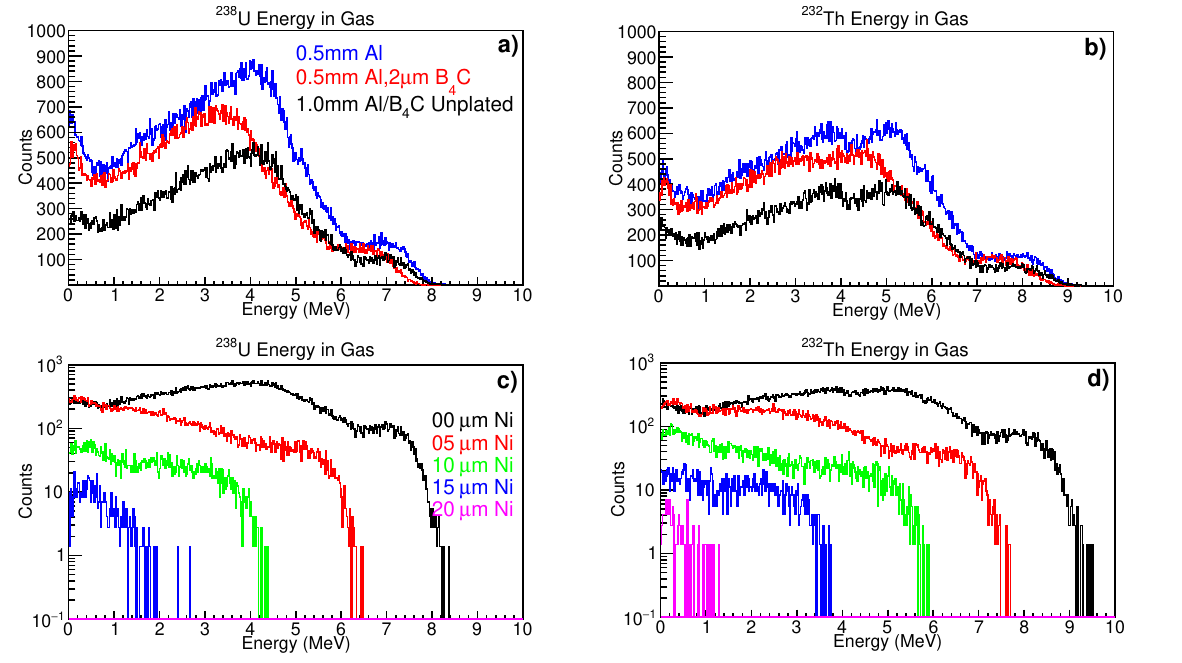}

\caption{\protect\label{fig:alpha1}Alpha energy-deposition distributions in
detector gas}
\end{figure}

Fig.~\ref{fig:alpha1}~a,b compare simulated gas-energy-loss distributions
after alpha emission from the 0.5~mm Al sheet, an Al sheet with $\mathrm{2\:\mu m}$
of $\mathrm{B_{4}C}$ evaporated on the surface and a 1.0~mm sheet
of Al-$\mathrm{B_{4}C}$ composite, which is 69\% Al by weight, for
$\mathrm{^{238}U}$ and $\mathrm{^{232}Th}$ decay sequences. The
plots have been scaled on the assumption that the actinide content
per unit volume of Al is constant. A coating of $\mathrm{2\:\mu m}$
$\mathrm{B_{4}C}$ on Al, as applied to rear normal blades of the
MG, reduces the alpha emission rate to about 75\% of the uncoated
sample. The emission rate from Al-$\mathrm{B_{4}C}$ (1~mm thick)
is somewhat lower than Al (0.5~mm thick) as relatively fewer alphas
make it to the surface of the thicker sample.

\subsection{\protect\label{subsec:Ni-Plating}NiP Plating of $\boldsymbol{\mathbf{\mathit{\mathrm{Al/B_{4}C}}}}$
Samples}

The measurements at the LBAS show that plating with NiP is highly
effective at stopping surface alpha emission from $\mathrm{Al/B_{4}C}$
sheets. However the NiP increases the probability of neutron scattering
\citep{MerlinTest}, so only the minimum plating thickness, consistent
with efficient stopping of alphas should be applied. 

Fig.~\ref{fig:alpha1}c,d display the dependence of gas energy loss
on the NiP-plating thickness applied to Al-$\mathrm{B_{4}C}$. The
simulation predicts that $\mathrm{25\:\mu m}$ of NiP will stop all
alphas, similar to the findings of Ref.~\citep{Birch}.

\begin{table}[H]
\begin{center}%
\begin{tabular}{|c|c|c|c|}
\hline 
NiP Thickness ($\mathrm{\mu m})$ & G4 $\mathrm{^{238}U}$ & $\mathrm{G4}\mathrm{\,{}^{232}Th}$ & Measured\tabularnewline
\hline 
\hline 
5 & $2.1\times10^{-1}$ & $3.1\times10^{-1}$ & --\tabularnewline
\hline 
10 & $2.9\times10^{-2}$ & $5.9\times10^{-2}$ & --\tabularnewline
\hline 
15 & $2.6\times10^{-4}$ & $1.1\times10^{-2}$ & --\tabularnewline
\hline 
20 & 0.0 & 0.0 & --\tabularnewline
\hline 
25 & 0.0 & 0.0 & $8.6\times10^{-4}$\tabularnewline
\hline 
\end{tabular}\end{center}

\caption{\protect\label{tab:Plating-ratio}Ratios of plated/unplated alpha
emission rate for different thickness of NiP plating on an Al-$\mathrm{B_{4}C}$
sample, for an alpha detection energy threshold of 1.5 MeV.}
\end{table}

\begin{figure}[H]
\begin{center}\includegraphics[width=0.5\columnwidth]{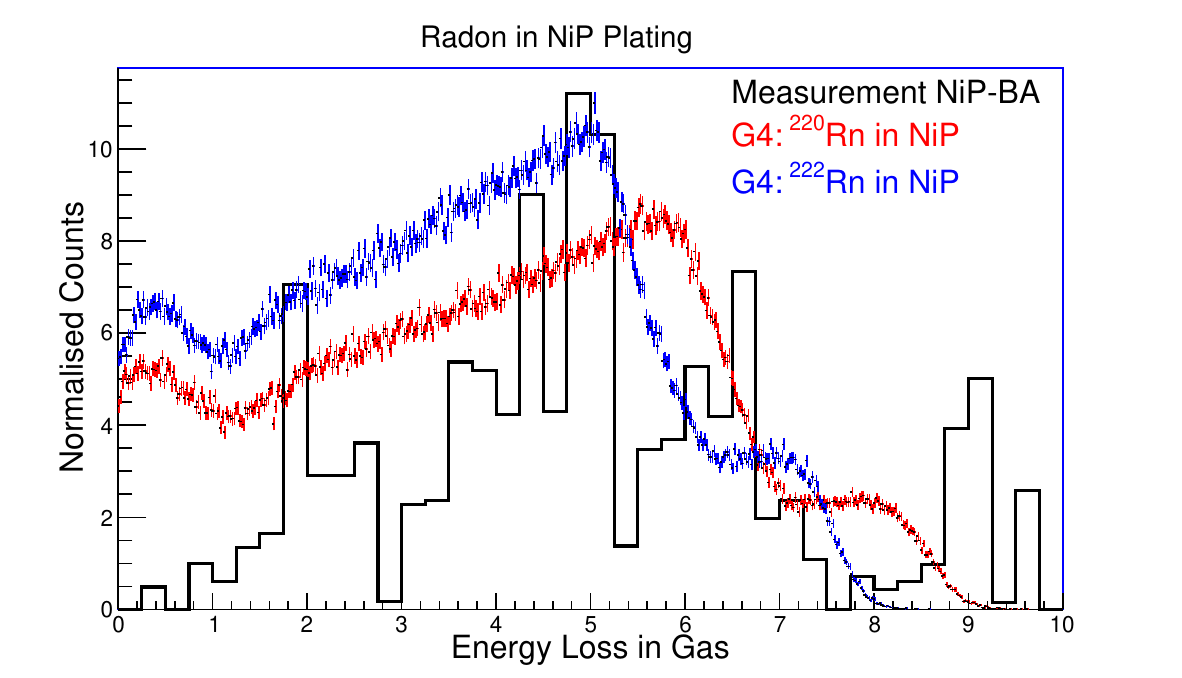}\end{center}

\caption{\protect\label{fig:Comparison-of-NiP-BA}Comparison of the NiP-BA
measurement and Geant4 simulations of the decay of radon isotopes
in the NiP plating. The Geant4 distributions have been scaled by 0.005
to fit on the measurement y-scale.}

\end{figure}

Sample NiP-BA has a nominal $\mathrm{25\:\mu m}$ thickness of NiP,
but shows a few counts even at 8-9 MeV (Fig.~\ref{fig:Alpha-pulse-height}),
which the simulation predicts should be absent even for 5-10 $\mu$m
of NiP plating. The sample was cleaned before measurement, but the
observed counts are consistent with close-to-surface deposition, if
the plating thickness is reasonably uniform. $\mathrm{^{220}Rn}$
and $\mathrm{^{222}Rn}$, gaseous daughter products of $\mathrm{^{232}Th}$
and $\mathrm{^{238}U}$, can diffuse through the sample before decaying,
leaving solid daughter products at the surface or inside the NiP plating.
Fig.~\ref{fig:Comparison-of-NiP-BA} compares the NiP-BA measurement
with a simulation of the decay of radon isotopes inside the $\mathrm{25\:\mu m}$
of NiP plating. The low statistics of the measurement limits any quantitative
comparison, but the simulation predicts that radon decay chains can
produce signals up to around 9~MeV.

\section{\protect\label{background}Background Counting Rate in T-REX Prototypes}

\begin{figure}[H]
\includegraphics[viewport=0bp 172.5bp 1055bp 542bp,clip,width=1\columnwidth]{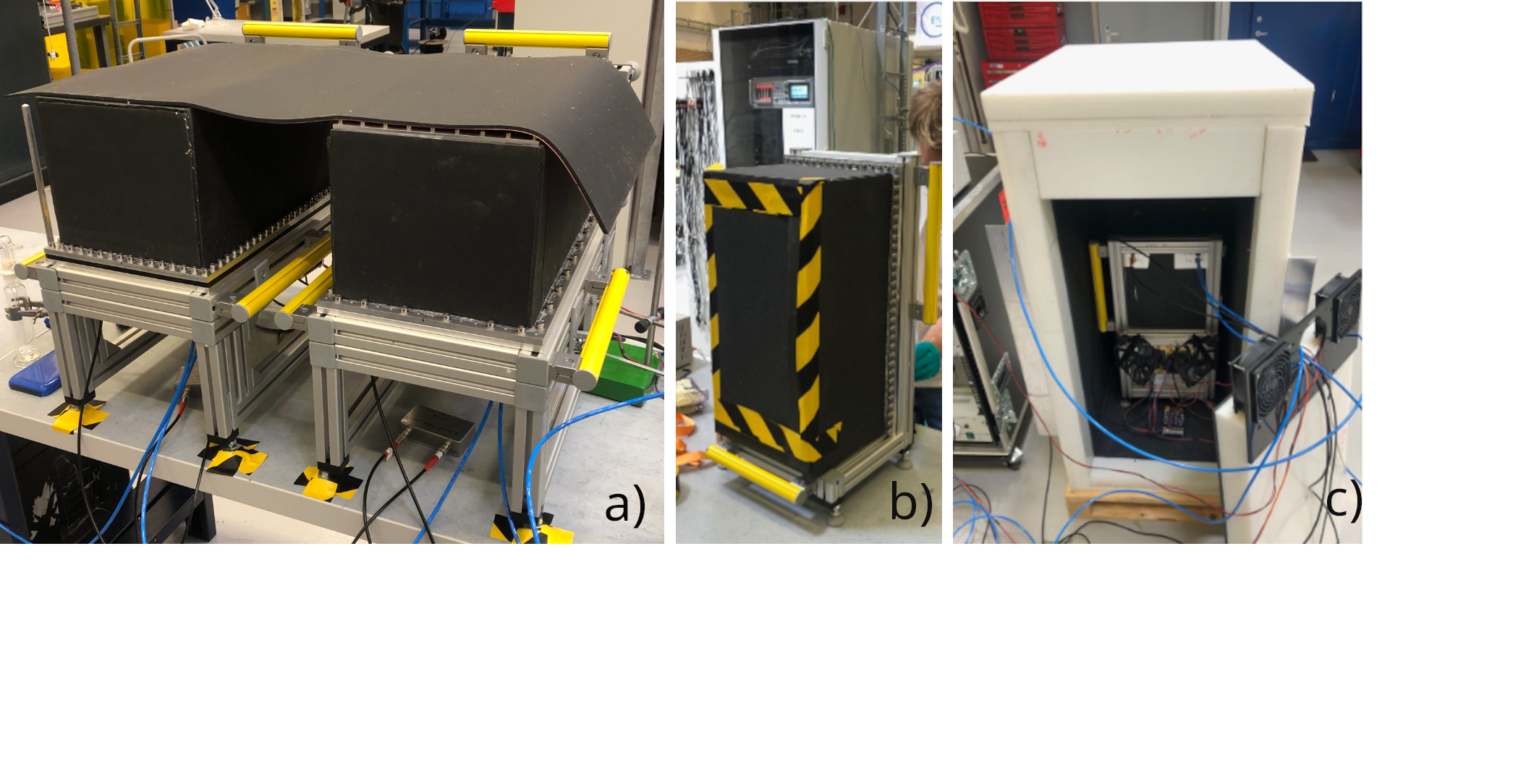}

\caption{\protect\label{fig:Detector-photo}MG background testing at ESS, a)
TRP-1 and TRP-3 in horizontal mode. A sheet of MB has been placed
over the front windows of the detectors. b) TRP-3 in vertical mode
with no outer shielding, c) TRP-3 in vertical mode with polyethylene
and MB outer shielding}
\end{figure}

Background rates have been measured for the two Multi-Grid prototypes
(Fig.~\ref{fig:Detector-photo}), TRP-1 and TRP-3 \citep{MerlinTest},
at ESS. TRP-1 has normal blades of 0.5~mm thick RP-Al, coated with
1-2~$\mathrm{\mu m}$ of $\mathrm{B_{4}C}$ and uncoated radial blades
of 0.5~mm RP-Al. TRP-3 has the same normal-blade configuration, but
the radial blades are 1~mm of Al/$\mathrm{B_{4}C}$, plated with
Ni. Two different plating techniques were tested, with the top 6 grids
using `electroless' NiP plating \citep{electroless} of the radial
blades, while the bottom 6 grids employed Ni electro plating.

The first background runs compared TRP-1 and TRP-3 rates, with the
detectors placed horizontally, while later runs concentrated on TRP-3
aligned vertically. The detector was optionally encased in polyethylene
shielding, lined with Mirrobor (MB\} \citep{Mirrobor}. In all cases
MB sheet was glued to the sides, tops and bottoms of the detectors

Both detectors employed VMM3A-based read out \citep{VMM}, with pulse-height
thresholds set to around 100~keV, which rejects events due to gamma
rays and also cosmic-ray muons. The neutron component of cosmic radiation
\citep{CosmicN,CosmicN1}, which peaks at $\sim1$~MeV and extends
from thermal to GeV energies, can trigger the MG and TRP-3 was tested
with and without MB-lined polyethylene shielding when in vertical
alignment. Although the maximum flux for the higher energy part of
the neutron spectrum ($E_{n}>20$~MeV) is expected at around vertical
incidence, lower energy neutrons will be more isotropic \citep{CosmicN1}.
Thus the orientation of a thermal neutron detector should not have
a strong effect on cosmic neutron rate.

\begin{figure}[H]
\includegraphics[width=1\columnwidth]{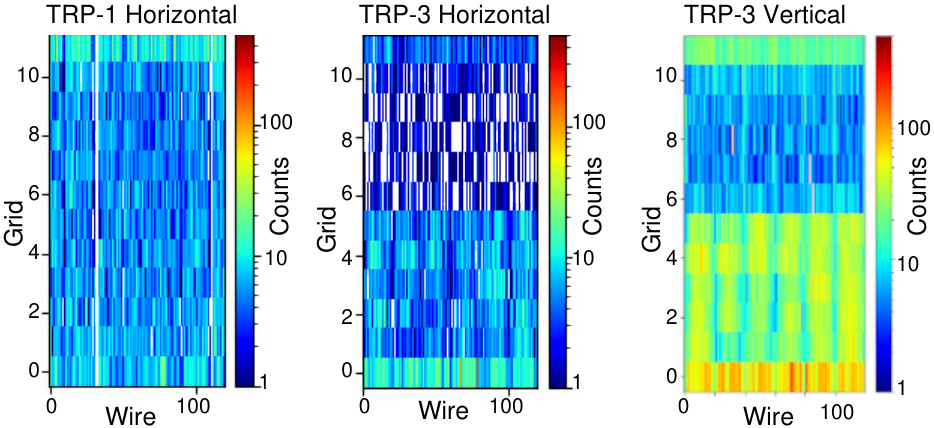}

\caption{\protect\label{fig:Background.}Background counts for each wire and
grid of the MG.}

\end{figure}

\begin{figure}[H]
\includegraphics[width=1\columnwidth]{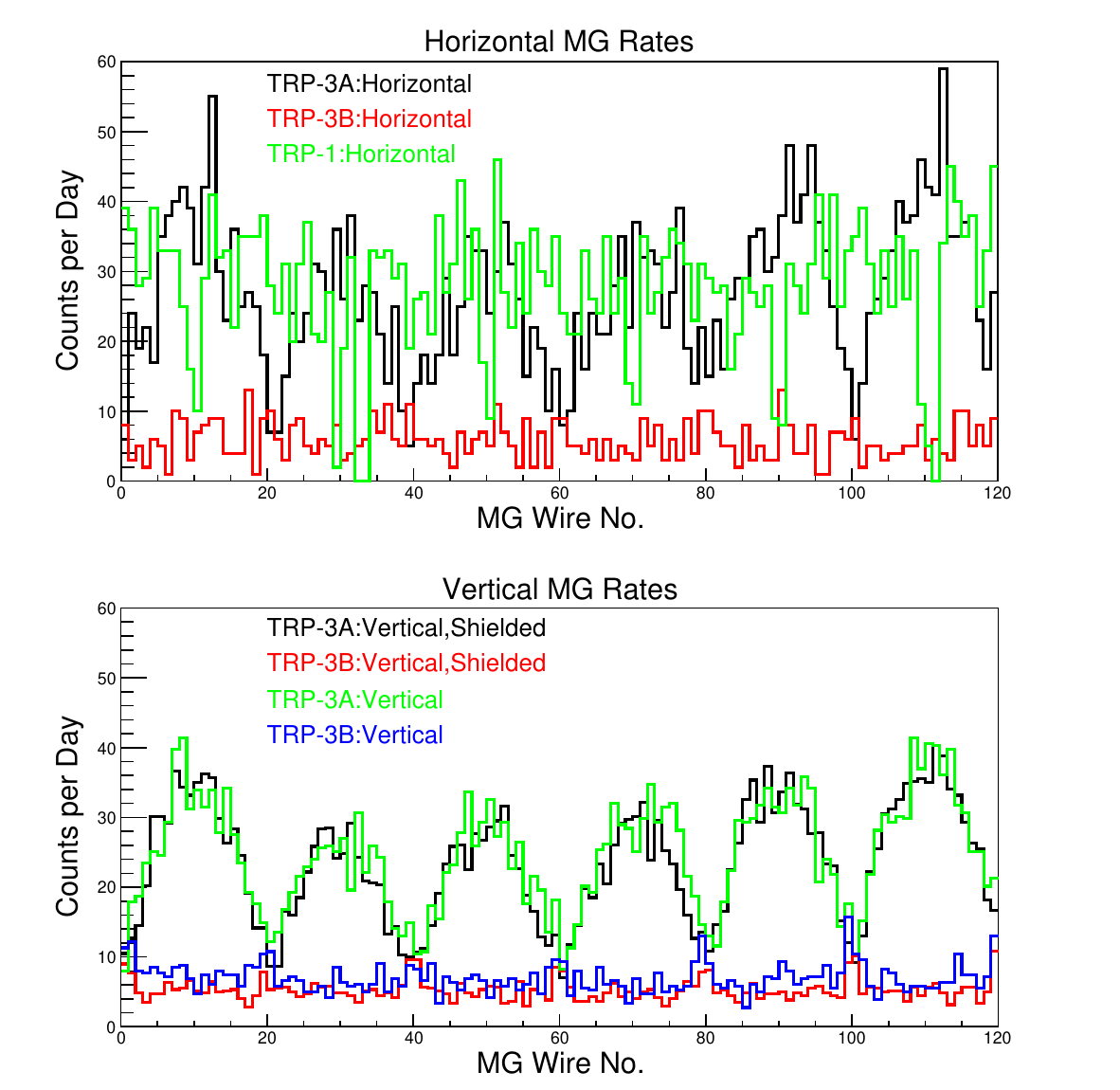}

\caption{\protect\label{fig:Background1}Wire counting rates for TRP-1 (summed
over grids 1-5), TRP-3A (grids 1-5) and TRP-3B (grids 6-10).}

\end{figure}

Fig.~\ref{fig:Background.} displays 2D plots of grid vs. wire counts
for runs with TRP-1 (left) and TRP-3 (centre) in horizontal alignment
and TRP-3 (right) in vertical alignment with extra polyethylene/MB
shielding. All plots show increased counting rate in bottom (0) and
top (11) grids. \textcolor{black}{This is attributed to the proximity
of non-radio-pure Al arms, which support the PCBs holding the MG wires
in tension (see Fig.~\ref{fig:Multigrid}), and also the Sn based
solder \citep{solder,Birch} which anchors the wire to the PCB. Alpha
emission from these components can impinge on the voxels of the top
and bottom grids so that grids 0 and 11 were excluded from further
counting rate comparisons. }Fig.~\ref{fig:Background.} \textcolor{black}{shows
that the upper TRP-3 voxels, with electroless radial-blade plating,
have lower background rates than those with electro plating in the
lower voxels.}

Fig.~\ref{fig:Background1} shows 1D plots of wire counting rates
for TRP-1, summed over grids 1-5 and TRP-3, summed over grids 1-5
(TRP-3A) and grids 6-10 (TRP-3B). The upper plot of rates in horizontal
alignment shows TRP-1 and TRP-3A \textcolor{black}{are very similar,}\textcolor{red}{{}
}while TRP-3B is significantly lower. The lower plot in vertical alignment
also shows that the TRP-3A rate is significantly higher than TRP-3B.
Shielded and unshielded configurations are quite similar. The TRP-3A
wire rates exhibit a cyclical structure with minimum rates at the
front and rear of the detector and maximum close to the centre. This
is attributed to a variation in electro-plating thickness along the
length of the radial blades, which minimises close to the centre of
the blade. 

Mean rates are given in Table~\ref{tab:Mean-counting-rates}. TRP-1
and TRP-3A horizontal are almost identical. Vertical rates in TRP-3A
are slightly lower than horizontal, while for TRP-3B the horizontal
rate is lower. Shielded rates are slightly lower than unshielded,
most noticeable in TRP-3B, where the overall rate is lower. 

Table~\ref{tab:Mean-counting-rates} also compares normalised rates
from the the TRP prototypes with two LBAS measurements and two detector
configurations from Ref.\citep{Birch}. The LBAS normalised rates
are considerably lower than those obtained from the prototypes and,
since the radial blades cover only 29\% of the total grid surface
area, one would expect the effect of the NiP-plated radial blade to
be relatively small. TRP-1 and TRP-3A rates are indeed similar, whereas
the TRP-3B rate is a factor $\sim5$ lower. From the background measurements
in Ref.\citep{Birch} prototype IN6, constructed from standard Al
alloy, showed a factor $\sim9$ higher rate than TRP-1 constructed
from RP-Al. Prototype P14 had a similar configuration to TRP-3B and
showed similar low rates.

\begin{figure}[H]
\includegraphics[width=1\columnwidth]{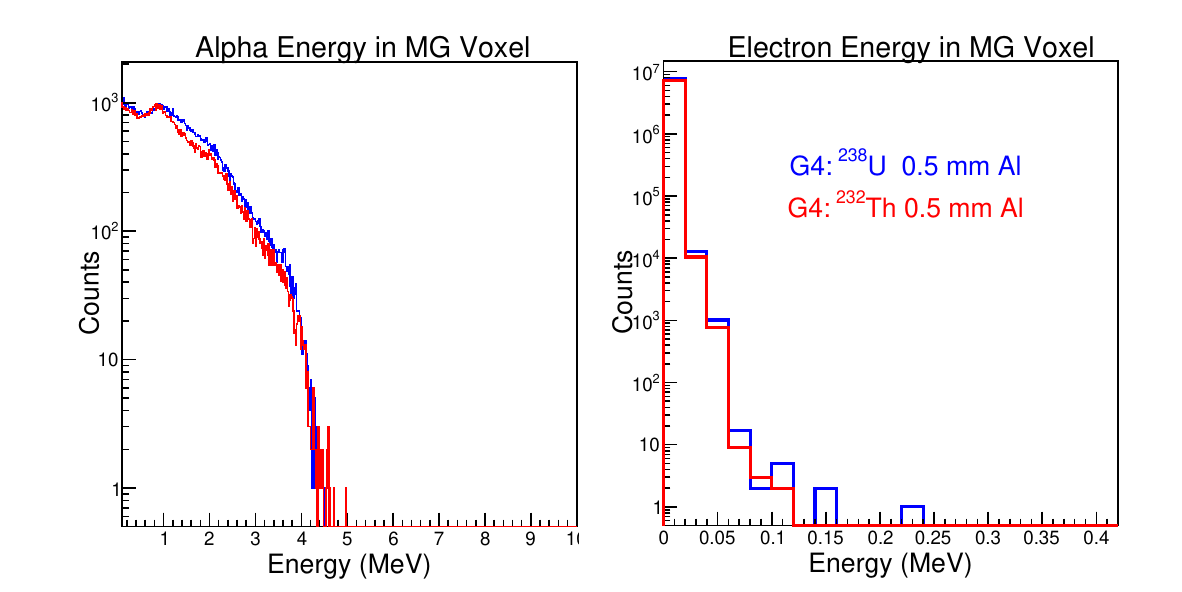}

\caption{\protect\label{fig:Simulated-Voxel}Simulated energy-loss distributions
of alphas and electrons produced by $\mathrm{^{238}U}$ and $\mathrm{^{232}Th}$
decay sequences, for a single MG voxel filled with a mixture of 80\%
Ar, 20\% $\mathrm{CO_{2}}$ at STP.}

\end{figure}

For the prototype measurements a detection threshold of 0.1 MeV \citep{VMM,MerlinTest}
was set to suppress background from gamma rays and cosmic muons, while
the LBAS measurements employed a threshold of 1.5~MeV. Geant-4 predicts
(Fig.~\ref{fig:Comparison-of-alpha}) the 1.5~MeV threshold removes
17\% of alphas and 100\% of electrons from the LBAS spectrum. Fig.~\ref{fig:Simulated-Voxel}
displays the simulated energy loss distributions in the Ar/CO2 mix
inside a single TRP-1 voxel defined by 0.5~mm Al normal and radial
blades. It is similar to an equivalent calculation in Ref.\citep{Birch}.
As the voxel has a smaller volume of gas than the LBAS there is less
material for the particles to traverse and smaller energy loss. A
threshold of 0.1 MeV removes 5.5\% of alpha events and almost 100\%
of electrons. Thus the simulation predicts that different energy thresholds
have only a modest effect on the relative alpha counting rates in
MG and LBAS.

However the different thresholds will affect the cosmic neutron background
quite differently, as the maximum energy of the products from thermal
neutron capture on $\mathrm{^{10}B}$ is only 1.4~MeV. A threshold
of 1.5 MeV removes all cosmics and, according to the simulation, 68\%
of decay alphas. Unfortunately the measured pulse height spectra from
the present prototypes have insufficient dynamic range for a meaningful
comparison. A comparison of TRP-3B rates with and without extra shielding
suggest some neutron cosmic ray contribution, but the effects of the
external shielding and the internal Al/$\mathrm{B_{4}C}$ radial-blade
shielding on TRP-3 remain to be quantified.

\begin{comment}
232Th 170500 168300 14220 84.5\%

238U 193800 190600 155200 81.4\%

Voxel surface area 79.2 cm2

IN6 rate 0.25 Hz/cm2/hr equivalent to 6Hz/cm2/day TRP equivalent 188.2

Factor 50 reduction 3.75
\end{comment}

\begin{table}[H]
\begin{center}%
\begin{tabular}{|c|c|c|c|c|c|}
\hline 
Detector/ & Alignment & Shielding & Time & Mean Rate & Normalised Rate\tabularnewline
Sample &  &  & (days) & ($\alpha$/day) & ($10^{-3}\alpha$/$\mathrm{cm^{2}}/$hr)\tabularnewline
\hline 
\hline 
TRP-1 & Horizontal & MB & 1.0 & $27.8\pm0.5$ & $14.6\pm0.3$\tabularnewline
\hline 
TRP-3A & Horizontal & MB & 1.0 & $27.04\pm0.5$ & $14.2\pm0.3$\tabularnewline
\hline 
TRP-3B & Horizontal & MB & 1.0 & $5.8\pm0.2$ & $3.1\pm0.1$\tabularnewline
\hline 
TRP-3A & Vertical & MB & 3.6 & $24.7\pm0.2$ & $13.0\pm0.1$\tabularnewline
\hline 
TRP-3B & Vertical & MB & 3.6 & $7.0\pm0.1$ & $3.7\pm0.1$\tabularnewline
\hline 
TRP-3A & Vertical & $\mathrm{C_{2}H_{4}}$+ MB & 6.0 & $23.6\pm0.2$ & $12.4\pm0.1$\tabularnewline
\hline 
TRP-3B & Vertical & $\mathrm{C_{2}H_{4}}$+ MB & 6.0 & $5.3\pm0.1$ & $2.8\pm0.1$\tabularnewline
\hline 
AL-2 (LBAS) & -- & -- & 13 & -- & 1.3$\pm0.1$\tabularnewline
\hline 
NiP-BA (LBAS) & -- & -- & 21 & -- & $0.3\pm0.1$\tabularnewline
\hline 
IL6 \citep{Birch} & -- & -- & -- & -- & 250\tabularnewline
\hline 
P14 \citep{Birch} & -- & -- & -- & -- & 4.7\tabularnewline
\hline 
\end{tabular}\end{center}

\caption{\protect\label{tab:Mean-counting-rates}Mean background counting rates.
In TRP prototypes MB sheet was placed over the outer detector surfaces.
Additional runs were made with TRP-3 placed inside a MB-lined polyethylene
enclosure ($\mathrm{C_{2}H_{4}}$+ MB). as shown in Fig.~\ref{fig:Detector-photo}c.
The normalised-rate entry for IL6 has been estimated from Fig.7 of
Ref.\citep{Birch}. From Table 2 of Ref.\citep{Birch} the P14 rate
was 1.86\% of IL6.}
\end{table}

\begin{figure}[H]

\begin{center}\includegraphics[width=0.85\columnwidth]{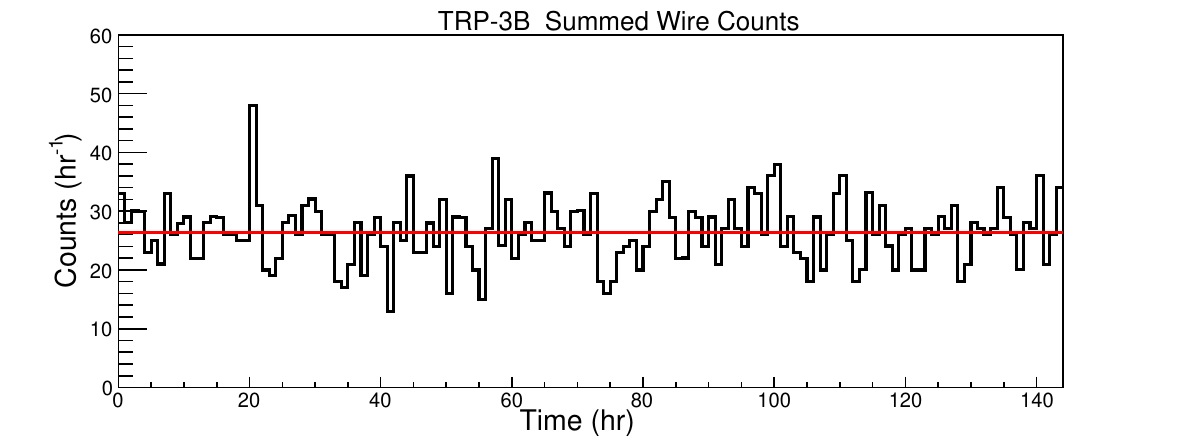}\end{center}\caption{\protect\label{fig:Time-dependence}The time dependence of the TRP-3B
background counting rate, summed over all 120 wires, during a 144~hr
run period. The mean rate ( $\mathrm{26.4\:hr^{-1}})$ is shown by
the horizontal red line. The statistical uncertainties range from
4 to 7 counts/hr.}

\end{figure}

The time dependence of the background from TRP-3B, measured during
the vertical-alignment runs with extra shielding installed, is displayed
in Fig.~\ref{fig:Time-dependence} and shows no statistically significant
departure from a constant rate over 144~hr of measurement. A constant
flat background should be straightforward to subtract when the spectrometer
operates in time-of-flight mode.

\section{\protect\label{subsec:Conclusions}Summary}

Multi-Grid technology, as employed in the T-REX neutron TOF spectrometer
at ESS, is fabricated in largely in Al. Al contains traces of actinides
(mainly descended from $\mathrm{^{232}Th}$ and $\mathrm{^{238}U}$)
which emit alpha particles, some of which escape into the active gas
volume of the MG producing a detectable signal and background counting
rate.

Alpha emission rates have been measured for 5 samples using a low-background
ionisation drift chamber. RP-Al, used for MG fabrication served as
a benchmark. Al-$\mathrm{B_{4}C}$ composite, a prospective material
for radial MG blades, showed a factor 280 increase in alpha rate,
compared to RP-Al. However a nominal $\mathrm{25\:\mu m}$ NiP plating
of the composite sample suppressed the alpha surface emission rate
to around 0.24 of RP-Al.

Geant-4 based simulations of surface emission from Al and composite
samples were performed to investigate the effect of coating Al or
composite sheets on the alpha emission rate. The calculated alpha
energy distributions after escaping an Al sample were qualitatively
similar to those measured. A 2~$\mu$m coating of $\mathrm{B_{4}C}$
on 0.5~mm RP-Al, as employed on the rear normal blades of the MG,
reduced the alpha emission rate by around 25\%. The effect of NiP
plating of Al-$\mathrm{B_{4}C}$ was investigated up to a thickness
of 25~$\mu$m, at which point no alphas escape. The measurement of
composite with a nominal 25~$\mu$m coating of NiP produced a small
but significant alpha counting rate and an energy spectrum extending
up to around 9~MeV. This is contrary to the simulation's prediction
if emission takes place inside the sample. However radon gas can diffuse
out of the sample interior into the NiP plating and beyond. Simulation
of $\mathrm{^{220}Rn}$ and $\mathrm{^{222}Rn}$ isotopes decaying
inside the NiP plating gives an alpha energy spectrum which displays
some of the features of the measurement.

Background rates were measured with 2 prototype MG detectors. TRP-1
has normal blades of RP-Al coated with 1-2 $\mathrm{\mu m}$ $\mathrm{B_{4}C}$
and radial blades of uncoated RP-Al, while TRP-3 has the same normal
blades as TRP-1, but the radial blades are Ni-plated Al-$\mathrm{B_{4}C}$.
Two types of Ni plating were compared: electroless plating with NiP
alloy and electro plating with Ni (on a thin deposit of Cu). Grids
with electroless plating had background rates on average a factor
4-5 lower that electro plated. The latter showed a systematic variation
in voxel rate along the length of the radial blades, peaking towards
the centre where the thickness of Ni was lower. Electroless NiP coating
should produce a more even thickness than electroplated Ni and indeed
the voxel rates with electroless radial blades show no significant
variation.

Background induced by the neutron component of cosmic radiation was
also tested for detector TRP-3. With extra polyethylene and MB neutron
shielding surrounding the detector, the mean rate on the electroless
plated voxels dropped from $7.0\pm0.1$ to $5.3\pm0.1$ counts/day.
However the effect of cosmic neutrons on the background rate remains
to be quantified and more study is necessary on the effectiveness
of neutron shielding. 

Based on these tests at ESS and other measurements at the ISIS spallation
source facility, the production of 88-grid columns for T-REX has started
using radial blades of 1~mm Al-$\mathrm{B_{4}C}$, electroless plated
with 25~$\mu$m of NiP, and normal blades of 0.5~mm RP-Al coated
with 1-2~$\mathrm{\mu m}$ of enriched $\mathrm{B_{4}C}$. Background
rates are being measured as part of the testing procedure of the columns
after assembly and will now be correlated with environmental factors
such as temperature and pressure.

\section*{Acknowledgements}

We wish to thank the technicians of the ESS Detector Group and the
University of Glasgow engaged in the design and construction of Multi-Grid
structures,

\textcolor{black}{The University of Glasgow acknowledge that the result
has been generated in collaboration with and through financial support
by European Spallation Source ERIC under Contract 325103. Glasgow
also acknowledge support from the UK Science and Technology Facilities
Council, Grant ST/V00106X/1.}

\end{document}